%% file: tada2024.tex
\documentclass{article}

    \PassOptionsToPackage{numbers, compress}{natbib}
\usepackage{amsmath}
\usepackage{adjustbox}
\usepackage[preprint]{neurips_2024}

\usepackage{colortbl}
\usepackage[utf8]{inputenc} % allow utf-8 input
\usepackage[T1]{fontenc}    % use 8-bit T1 fonts
\usepackage{hyperref}       % hyperlinks
\usepackage{url}            % simple URL typesetting
\usepackage{booktabs}       % professional-quality tables
\usepackage{amsfonts}       % blackboard math symbols
\usepackage{nicefrac}       % compact symbols for 1/2, etc.
\usepackage{microtype}      % microtypography
\usepackage[dvipsnames]{xcolor}         % colors
\usepackage{graphicx}

\definecolor{mygreen}{HTML}{cbf078}

\title{Blind Data Adaptation to tackle Covariate Shift\\ in Operational Steganalysis}

\author{%
  Rony~Abecidan \\
    Univ. Lille, CNRS, Centrale Lille\\
  UMR 9189 CRIStAL\\
  F-59000 Lille, France \\
  \texttt{ronyabecidan@protonmail.com} \\
  \And
  Vincent~Itier \\
  Centre for Digital Systems, Univ. Lille, CNRS, Centrale Lille,\\
  UMR 9189 CRIStAL\\
  F-59000 Lille, France \\
  \texttt{vincent.itier@imt-nord-europe.fr} \\
  \And
  Jérémie~Boulanger \\
  Univ. Lille, CNRS, Centrale Lille,\\
  UMR 9189 CRIStAL\\
  F-59000 Lille, France \\
  \texttt{jeremie.boulanger@univ-lille.fr}
  \\
  \And
  Patrick~Bas \\
  Univ. Lille, CNRS, Centrale Lille,\\
  UMR 9189 CRIStAL\\
  F-59000 Lille, France \\
  \texttt{patrick.bas@cnrs.fr}
  \\
  \And
  Tomáš ~Pevný \\
  Department of Computers and Engineering\\
  Czech Technical University\\
  Prague, Czech Republic \\
  \texttt{pevnak@protonmail.ch}
}

\begin{document}

\maketitle

\begin{abstract}
The proliferation of image manipulation for unethical purposes poses significant challenges in social networks. One particularly concerning method is Image Steganography, allowing individuals  to hide illegal information in digital images without arousing suspicions. Such a technique pose severe security risks, making it crucial to develop effective steganalysis methods enabling to detect manipulated images for clandestine communications. Although significant advancements have been achieved with machine learning models, a critical issue remains: the disparity between the controlled datasets used to train steganalysis models against real-world datasets of forensic practitioners, undermining severely the practical effectiveness of standardized steganalysis models. In this paper, we address this issue focusing on a realistic scenario where practitioners lack crucial information about the limited target set of images under analysis, including details about their development process and even whereas it contains manipulated images or not. By leveraging geometric alignment and distribution matching of source and target residuals, we develop $\mathtt{TADA}$ (Target Alignment through Data Adaptation), a novel methodology enabling to emulate sources aligned with specific targets in steganalysis, which is also relevant for highly unbalanced targets. The emulator is represented by a light convolutional network trained to align distributions of image residuals. Experimental validation demonstrates the potential of our strategy over traditional methods fighting the steganalysis covariate shifts.
\end{abstract}

\input{1_introduction.tex}

\input{2_formalization.tex}

\input{3_training.tex}

\input{4_experiments.tex}

\bibliographystyle{ieeetr}
\bibliography{refs}

%%%%%%%%%%%%%%%%%%%%%%%%%%%%%%%%%%%%%%%%%%%%%%%%%%%%%%%%%%%%

\appendix

\input{5_appendix.tex}

\end{document}

%% file: 1_introduction.tex
\section{Introduction}

\textit{Steganography}, the science of concealing information within seemingly innocuous data, is today a formidable tool for cybercriminals. With the help of numerous softwares available online, it is indeed possible for anyone to hide secret information into text, videos and images, enabling therefore to communicate secretly or embed viruses in total discretion \cite{stegaattack}. Such security concerns justify the popularity of steganography in the literature. Among the most famous steganographic schemes, those tailored for images are particularly studied. This is explained by the significant embedding capacity afforded by digital images combined with their widespread distribution online.
\begin{figure}
    \centerline{
    \includegraphics[width=\linewidth]{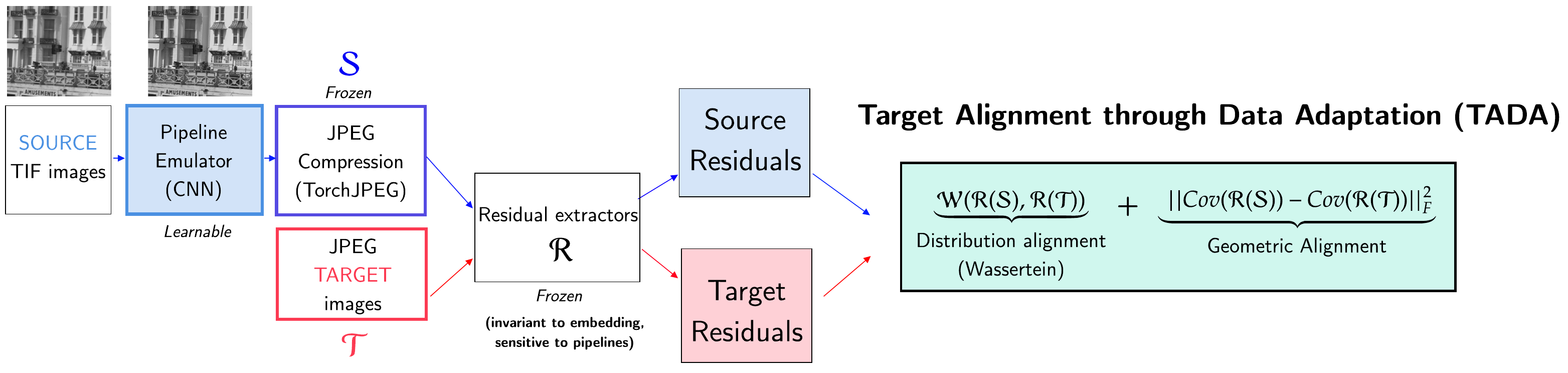}}
    \caption{Illustration of Target Alignment through Data Adaptation ($\mathtt{TADA}$). Data adaptation consists in learning a small convolutional network which emulates the development pipeline. The alignment is performed by using a dual loss matching the distribution of the residuals of the target images with the distribution of the residuals of the emulated image coming from the source.
    $\mathtt{TADA}$ proposes an end-to-end emulation mechanism to realign the two distributions in order to reduce covariate shift. }
    \label{fig:tada}
\end{figure}
% Lower part: we highlight using scatter plots of image residuals, the fact that the covariate shift in steganalysis is due a discrepancy between the covariance matrices of the source and target residuals (see also section~\ref{sec:analysis}). Each point represents two neighboring samples of an image subject to a high-pass filter, {\it i.e.} a 2D residual. 
% As example, methods such as HUGO \cite{hugo} and HILL \cite{hill} are renowned for their effectiveness in embedding messages into the spatial domain of digital images. 
The JPEG domain is for instance well studied for steganography due to its prevalent use on social networks. Among the well-known embedding techniques within this domain, we have UERD \cite{uerd} and J-UNIWARD \cite{juniward}, which are respectively based on DCT and wavelet decompositions. These techniques leverage textured areas of images to embed messages within their natural noise, thereby making detection more arduous. 
Forensic analysts are aware of these manipulations and with the help of machine learning, they train \textit{Steganalysis} detectors to determine the authenticity of images under scrutiny in the context of criminal investigations.

Because steganographic manipulations are imperceptible and can manifest in diverse forms, it is common to leverage supervised models to build steganalysis detectors. It requires therefore the use of a training base of genuine (a.k.a \textit{covers}) and manipulated (a.k.a. \textit{stegos}) images to build steganalysis detectors. Researchers are nowadays training their steganalysis detectors using toy cover images coming from carefully prepared databases like BOSSBASE~\cite{bossbase} or ALASKABASE~\cite{alaskabase}. 

It has been shown in~\cite{spam,srm,dctr} that noise residuals are helping to extract discriminative features for Steganalysis. Provided the payload of the images under scrutiny is large enough\footnote{This notion of a "large" payload is relative to the noisy properties of the images at stake}, a simple linear classifier can leverage these features to perform efficient steganalysis. Naturally, high levels of noise within an image facilitate the concealment of a message, thereby making it more challenging to differentiate between covers and stegos.
For the most difficult cases, the state of the art for steganalysis lies on convolutional neural networks such as YedroudjNet \cite{Yedroudj}, SRNet \cite{SRNet} and JIN-SRNet \cite{pretrainforstega} that are also using noise residuals by training several high resolution convolutional layers to improve their predictions.
Unfortunately for practitioners, steganalysis detectors are naturally failing with real-world images that come from completely unknown distributions. In machine learning, this scenario is well-known and referred as \textit{covariate shift} in \cite{datashift} while in steganalysis, this is referred as \textit{Cover-Source Mismatch (CSM)} since it results directly from a mismatch of cover distributions~\cite{giboulot,csm_chaumont,csm_fridrich}.
This mismatch comes from the fact that significant variation is found in cover distributions, commonly known as cover sources, which is due to several aspects of the image acquisition process. These aspects range from the type of capturing sensor (such as CCD or CMOS), its quality (its inherent ability for a given ISO setting to generate a photonic noise of small power), the capturing parameters (including ISO, exposure time and aperture), the traits of the captured image (like lighting conditions and content uniformity), the post-processing actions (such as white balance adjustment and gamma correction), to the compression steps that follow (such as 8-bit conversion and JPEG compression).
% Each operation of a processing pipeline involves parameters that shape the statistical properties of the generated covers to any steganographic embedding.
These operations used for visual enhancing and compression are modifying the noise distribution upon which steganalysis detectors rely. Thus, given the sensitivity of machine learning models to their training distributions, the processing pipeline is broadly identified as the mainstay of CSM~\cite{giboulot}.

\section{Contributions}
\label{s:contri}

While several strategies are proposed to address covariate shift in machine learning (\cite{da1,da2,da3,da4,da5,deepda1,domainrandomization}) and CSM in steganalysis (\cite{csm_chaumont,csm_fridrich,giboulot,wifs2022,wifs2023}), very few of them are really efficient in the realistic scenarios where:

\begin{itemize}
    \item The processing pipeline of images under scrutiny is totally unknown (neither the nature of the operations undertaken nor the hyperparameters used for each operation are known). 
    \item The steganalyst observes only a small set of images with unknown development pipelines.
    \item We are ignorant about the class of each image to analyze (cover or stego). There is possibly a high unbalance in terms of cover-stego image among the test set (extreme unbalance appears most of the time because in practice most users are innocent). 
    
    % Consequently, we cannot rely on supervised alignment methods and we need to use features which are invariant to the embedding.
\end{itemize}

Faced with that reality, our contributions are twofold:

\begin{itemize}
    \item We propose a novel domain adaptation strategy to cope with covariate shift in JPEG Steganalysis based on a new geometric interpretation of CSM illustrated in Figure~\ref{fig:tada}. This strategy lies on $\mathtt{TADA}$ (Target Alignment through Data Adaptation), a small convolutive architecture learning how to develop diverse RAW images from ALASKABASE \cite{alaskabase} so that noise residuals of target and source images are aligned with the ones of the images under scrutiny. 
    \item This data adaptation strategy relies on a combination of two complementary losses. The first term considers the alignment of principal axis (eigenvectors) and spreading (eigenvalues) of both residual distributions equalizing covariance matrices. The second term considers the Wassertein distance between these same distributions in order to be robust to bias which are not captured by the first one.% in both geometry and distributions.
    %\item We offer a fresh geometric perspective on CSM in Steganalysis illustrated in Figure~\ref{fig:cov_alignment}. This perpective both allows assessing the reliability of a detector trained on a specific dataset when tested on scrutinized images, but also explains why the non-alignment of the covariance of the image residuals is the main cause of Covariate shift/CSM in steganalysis. 
\end{itemize}

% We strongly believe that approximating accurately the distribution of original images is pivotal for creating steganalysis detectors tailored to the needs of forensic analysts. Hence, we designed $\mathtt{TADA}$
%  to achieve that goal assuming two reasonable hypothesis on scrutinized JPEG images: 

%  \begin{itemize}
%     \item The processing pipeline of images under scrutiny is totally unknown (Neither the nature of the operations undertaken nor the hyperparameters used for each operation are known). 
%     \item Only few of these images are available.
%     \item We don't have a clue about their nature (cover or stego).
%     \item Extreme unbalance frequently appears (in most cases users are innocent).
% \end{itemize}

As far as we know, this data adaptation strategy is the first effort to address covariate shift in steganalysis by proposing a neural architecture designed to emulate a relevant source dataset with desired target statistics, especially in cases where our knowledge about these targets is very limited. Both toy and real-world experiments underscore the potential of $\mathtt{TADA}$ over state of the art methods available to forensic practitioners (see Section~\ref{sec:experiments} and Appendix \ref{app:ablations}). In contrast with other strategies such as training on appropriate mixtures of sources or, picking the closest source from a relevant set of sources, our method is adaptive and enable to emulate a pipeline as close as possible to the true development pipeline used to generate the test images.

\subsection{Outline of the paper}

Section~\ref{sec:relatedworks} discusses related research on covariate shift in steganalysis from both machine learning and steganalysis perspectives.
In section~\ref{sec:analysis}, we introduce the $\mathtt{TADA}$ approach, starting with the formalization of our steganalysis issue followed by a detailed explanation of the learning mechanism designed to emulate the development pipeline of a specific target. Afterwards, Section~\ref{sec:trainingdetails} presents an overview of $\mathtt{TADA}$ training, detailing important pre-processing steps and the learning metrics that are optimized. Finally, Section~\ref{sec:experiments} demonstrates through experiments that $\mathtt{TADA}$ can surpass state-of-the-art methods on both toy and real-world targets, even when the targets are unbalanced and contain a limited number of images.

% we strongly believe that approximating accurately the distribution of original images is pivotal for creating steganalysis detectors tailored to the needs of forensic analysts. 

% very important for steganalysis and more generally, image forensics in the wild. If you know to match the sources, you can 

% : The task to perform is the same, the stego generation process given inputs remain unchanged and, the input distributions are differing

% Following Kerckhoffs' principle, we assume that Alice embedding scheme is known, while her secret key remains undisclosed. 

% Knowing the cover source is important for steganography and image forensics in the wild (cite forensics sources). This the motivation : if you know to match the sources, you can improve the detector.

% To be as realistic as possible, we assume that we can only gather a small amount of unlabeled images $(<=500)$ without having any clue about the way they have been created. We will also consider that we don't know the proportion of manipulated images in this set, acknowledging the possibility to have 100\% covers or 100\% stegos.

% add figure of image creation
\section{Related Works}\label{sec:relatedworks}

% add figure of scenario

Strategies commonly employed to address covariate shift in a blind scenario can be categorized into two frameworks. 

First, there are  \textit{Domain Randomization} strategies used for instance in~\cite{domainrandomization} and~\cite{domainrandomization2} aiming at building a training set fostering noise and content diversity through a mixture of multiple image distributions. This essentially consists in augmenting the training set in a clever way so that we  achieve broad generalization across various targets. 
% ALASKABASE~\cite{alaskabase} is a good example of this approach by compiling 80k images from different sensors, captured at varying ISO levels.
Although~\cite{holisticker} demonstrates the potential of this strategy against CSM, reference~\cite{wifs2022} suggests that not all combinations of distributions are equally effective, emphasizing that quality prevails over quantity for optimal generalization. Thus, maximizing the generalization ability of a steganalysis detector requires identifying the most suitable combination of distributions for a specific testing set. Studies like~\cite{csm_fridrich} and~\cite{wifs2023} are precisely proposing metrics to assess the relevance of a cover source w.r.t. a specific target. 

% If we have access to numerous images of the target set, we can correctly estimate these metrics.  Under the assumption that we have sources that match the target set, such metrics are particularly relevant for target generalization.

Secondly, \textit{Unsupervised Domain Adaptation} strategies \cite{uda} make the most of all available data to tackle covariate shift. In this scenario, the goal is to transfer knowledge obtained from a labeled training set (a.k.a the \textit{source}) to an unlabeled evaluation set (a.k.a. the \textit{target}). This problem is well-known in machine learning and one famous strategy to cope with it involves embedding source and target into a common domain invariant space, so that the domain discrepancy term of the generalization bound of Ben David {\it et al.} \cite{gen_bound} is minimized.
For instance, Ganin {\it et al.}~\cite{deepda2} harness backpropagation to directly learn a domain-invariant projection while training a classifier. Their approach entails integrating and adversarially training a domain discriminator into the final layer of the neural network, hence fostering the creation of a representation where distinguishing between source and target domains becomes challenging. In the same vein, Long {\it et al.} introduce in 2015 a Deep Adaptation Network (DAN)~\cite{deepda1} allowing feature transferability in downstream layers of a CNN, ensuring that source and target distributions are close in the last projections through the minimization of the Maximum Mean Discrepancy (MMD).

At last, there are also strategies aiming to find a transformation directly embedding the source into the target domain.
For  example,  Fernando et al. propose in~\cite{da2} to align the subspaces spanned by source and target eigenvectors after applying a PCA. This strategy is very similar to the one of~\cite{wifs2023} where the authors propose to deduce the processing pipeline applied to scrutinized images minimizing the chordal distance between source and target DCTr features~\cite{dctr} with a simulated annealing. Expanding upon this idea, Sun et al. suggest in~\cite{da3} to recolor whitened source features with the covariance of the target distribution. By essence, this covariance alignment encapsulates the one of~\cite{da2} and~\cite{wifs2023} since the PCA projections are directly derived from covariances. 

Although all these strategies have proven fruitful in several experiments, they have obvious limitations to tackle the practical setup introduced in Section~\ref{s:contri}:
\begin{itemize}

    \item Many of them are relying on distances between distributions that are difficult to approximate with very few samples although their low sample complexities~\cite{sampling_complexity1,sampling_complexity2}.

    \item They are likely uneffective in the realistic case of highly unbalanced targets~\cite{unbalancedda}.

    \item In cases where source and target are balanced, nothing prevent the distribution to match while inverting the classes as illustrated in Figure 1 of~\cite{unbalancedda}. 
\end{itemize}

% \subsection{Why the covariance matrix is invariant to stego embedding?}

% For this approach we rely on the covariance matrix to extract a reliable fingerprint of the development pipeline which is also, just as importantly, largely invariant to the steganographic embedding. This last property is due to the fact that already early embedding schemes such as Model-Based steganography~\cite{sallee2003model} or nsF5~\cite{fridrich2007statistically} designed their embedding in order to preserve at least the second order moment of the Cover distribution, or the whole marginal distribution during the embedding. As a result, the Covariance matrix is well preserved but at the same time represents a strong fingerprint of the development pipeline, this observation was also experimentally verified in~\cite{mallet2024statistical}.    

%% file: 2_formalization.tex
\section{Approach}\label{sec:analysis}

In this section, we present $\mathtt{TADA}$ (Target Alignment through Data Adaptation), a new strategy to derive relevant sources for specific targets in steganalysis. The main idea is to learn how to process images with minimalist development in order to match target statistics. $\mathtt{TADA}$ is made of three key ingredients : a \textbf{pipeline emulator}, followed by a \textbf{feature extractor} sensitive to the processing pipeline while robust to steganographic embedding and, a \textbf{two-objectives loss function} aligning these features in terms of geometry and distributions. These steps are illustrated in Figure~\ref{fig:tada} which present an overview of the method. 
% multi objective instead of \textbf{dual-loss function}
\subsection{Formalization}

Using the notations from \cite{csmforma}, we describe a processing pipeline as a vector $\omega \in \Omega$ (the infinite set of possible processing pipelines), which includes parameters like the denoising coefficient and JPEG quality factor. In steganalysis, we add an extra parameter $\gamma$ to represent choices made by the steganographer, such as the embedding method and payload. Our task is to develop an algorithm able to distinguish covers from stegos among a set of images. Machine learning models are typically employed for this task:
\begin{align*}
    f(x \mid \theta_{\omega,\gamma}) : \ & \mathcal{X} \rightarrow \{cover,stego\}. \\
    & x \mapsto y
\end{align*}
where $\theta_{\omega, \gamma} \in \Theta$ represents all the parameters learnt using covers issued from the pipeline $\omega$ and potentially embedded following $\gamma$.

We consider now the unsupervised domain adaptation framework. We assume having access to $n_s$ labeled i.i.d. images ${\mathcal S}=\left\{x_i^s, y_i^s\right\}_{i=1}^{n_s}$ from $p((x,y) \vert \omega^{\mathcal S},\gamma^{\mathcal S})$ and $n_t$ unlabeled i.i.d. images ${\mathcal T}=\left\{x_i^t\right\}_{i=1}^{n_t}$ from $p(x\vert\omega^{\mathcal T},\gamma^{\mathcal T})$ with  $n_t<<n_s$. We aim to minimize the risk of failure of a detector trained on source data and evaluated on target data, such as: %We aim to minimize the risk of failure when we train a detector with source data and evaluate it with target data :
\[\mathbb E_{(x,y) \sim p((x,y) \vert \omega^{\mathcal T},\gamma^{\mathcal T})} (f(x \mid \theta_{\omega^{\mathcal S},\gamma^{\mathcal S}}) \neq y).\]
Following Kerckhoffs' principle, we assume that $\gamma^{\mathcal T}$ is known. It is therefore possible for practitioners to reproduce the embedding strategy having $\gamma^{\mathcal S} = \gamma^{\mathcal T} = \gamma$. However, we assume $\omega^{\mathcal S} \neq \omega^{\mathcal T}$ leading forensic analysts to a mismatch of cover distributions causing a covariate shift \cite{datashift} : $p(x|\omega^{\mathcal S}) \neq p(x|\omega^{\mathcal T})$ while $p(y|x^s,\gamma) = p(y|x^t,\gamma)$. This mismatch can be directly resolved if we manage to bring $\omega^{\mathcal S}$ as close as possible to $\omega^{\mathcal T}$. We will assume here that we can at least get access to the quantification tables of the target images\footnote{This information is often public. It's also possible to estimate it \cite{qfestimation}}.

\subsection{Emulation of realistic pipelines}

Among all the possible operations we could perform on images before JPEG compression, we show in \cite{wifs2022} that denoising and  sharpening are highly responsible of the covariate shift observed in steganalysis. Although these operations are not always linear, there exist linear versions of them such as Median Filter or Unsharp Masking largely used in classical softwares like Photoshop, GIMP, RawTherapee, \emph{etc}. More precisely, traditional image operations can be reasonably approximated with symmetric convolutions which sums to 1 as assumed in \cite{farid}. %For instance, a kernel of size $3 \times 3$ satisfying these conditions are following the shape :
For instance,  $3 \times 3$ kernel satisfying these conditions are structured as:

\begin{equation}
    \label{eq:kb}
    \begin{bmatrix}
        b & c&b \\
        c & a & c \\
        b &c & b
    \end{bmatrix}
    \text{ with  } a+4(b+c)=1.
\end{equation}

We propose therefore to use a unique convolution of this shape to emulate the target pipeline in $\mathtt{TADA}$. By choosing such a simple developer, we cannot reasonably approximate highly non-linear operations as well as resizing operations. We are aware of this limitation and we are working on making $\mathtt{TADA}$ compatible with such operations. However, we show here that a simple constrained convolution can already be very helpful in practical situations.

To be able to learn this convolution appropriately, we define now a differentiable metric assessing the proximity of two developping pipelines given the images they produce.

\subsection{Derive processing pipelines fingerprints from noise residuals}%with 

By nature steganalysis focuses on weak signals that do not rely on the image content. Camera sensors introduce different types of noise into captured images~\cite{noise}. In the RAW format, immediately after acquisition, the noise at the pixel level is independently distributed and follow a Poisson/Gaussian distribution~\cite{noiseraw}. Once a processing pipeline is applied to these RAW images, the noise pixels are correlated. 

% \vspace*{-5cm}
Since different processing pipelines affects noise correlations differently, Mallet {\it et al.}~\cite{mallet} propose therefore to consider these noise correlations as fingerprints of the processing pipeline. This approach is especially interesting in steganalysis considering the high sensitivity of noise residuals to the development pipeline, but also considering the robustness of theses fingerprints to the embedding, which is experimentally verified in~\cite{mallet}. We can explain this robustness by considering that even early embedding schemes such as Model-Based steganography~\cite{sallee2003model} or nsF5~\cite{fridrich2007statistically}, designed their embedding in order to preserve at least the second order moment of the Cover distribution, or the whole marginal distribution during the embedding. 

\begin{figure}[h!]
    \centerline{
    \includegraphics[width=0.9\linewidth]{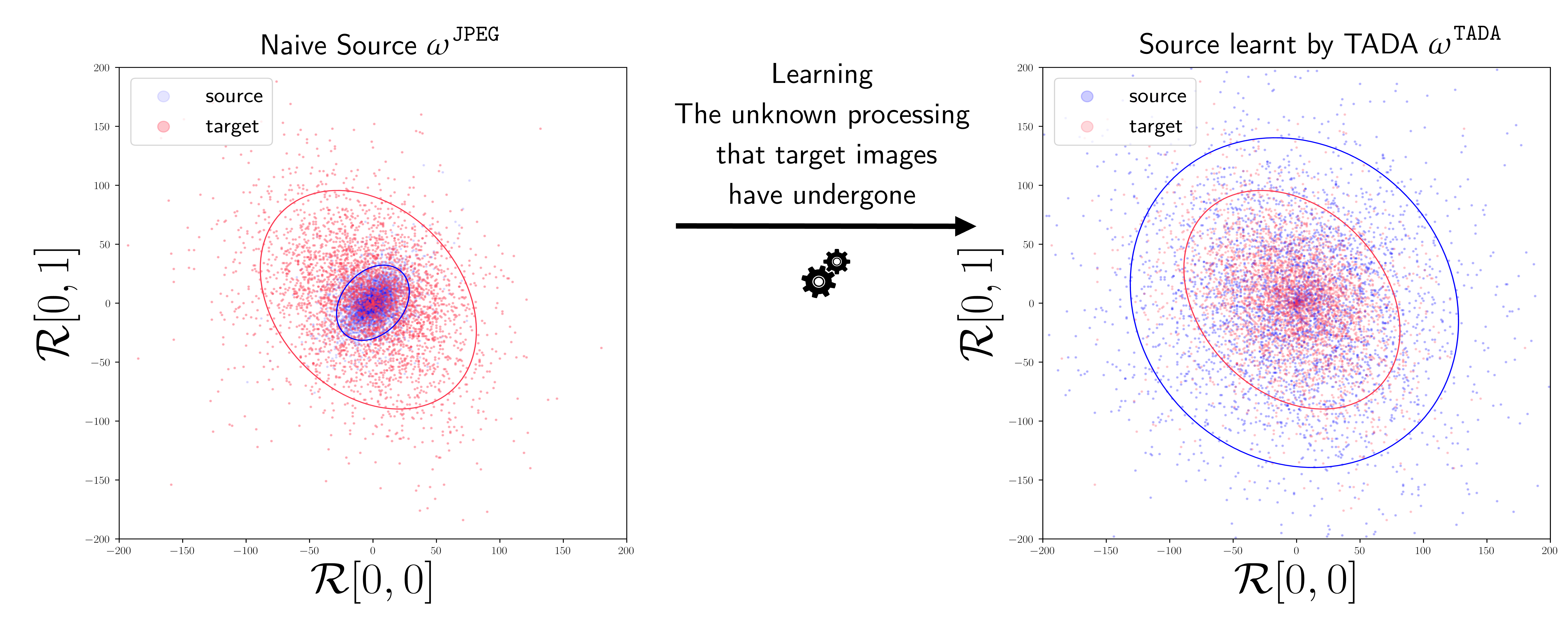}}
    \caption{Covariance alignment performed by $\mathtt{TADA}$ - We highlight using scatter plots of image residuals, the fact that the covariate shift in steganalysis may be caused by a discrepancy between the covariance matrices of source and target residuals. Each point represents two neighboring samples of an image subject to a high-pass filter $\mathcal R$, {\it i.e.} a 2D residual. }
    \label{fig:cov_alignment}
\end{figure}

A simple approach to extract noise residuals involves using high-pass filters on the images \cite{srm}. There are several filters designed for residuals extraction removing steganalysis traces while highlighting inter-pixel correlations, such as the KB filter~\cite{kbfilter} with coefficients $[[-\frac{1}{4},\frac{1}{2},-\frac{1}{4}] [\frac{1}{2},-1,\frac{1}{2}] [-\frac{1}{4},\frac{1}{2},-\frac{1}{4}]]$.

% \begin{equation}
%     \label{eq:kb}
%     \begin{bmatrix}
%         -0.25 & 0.5 &-0.25 \\
%         0.5 & -1 & 0.5 \\
%       -0.25 &0.5 & -0.25
%     \end{bmatrix}.
% \end{equation}

Referring to $\mathcal{R}$ as a high-pass filter capturing image noise residuals, \cite{mallet} inspires us to leverage $||\operatorname{Corr}(\mathcal R(\mathcal{S}))-\operatorname{Corr}(\mathcal R(\mathcal{T}))||_{F}$ as a differentiable loss function to guide the learning of $\omega^{\mathtt{TADA}}$. However, after investigations we discovered that equalizing noise correlations is not enough to avoid Cover Source Mismatch. We explain this observation by scale and shift invariance properties of correlations, leading our pipeline learning towards suboptimal pipelines. Hence, rather than equalizing correlations, we propose to equalize covariances to preserve both alignment (e.g. eigenvectors) and spreading (e.g. eigenvalues) of noise residuals.   

This approach is based on a key observation: directions with high variance in noise residuals are ideal for steganographers to hide their messages. Consequently, if there are specific directions in the residual distribution where the variance is high, steganographic methods will push the distribution in these directions. To distinguish between stegos and covers, we can project all images onto these high-variance directions and apply a variance threshold. Stego images will show significantly higher variance compared to cover.
However, if the source pipeline differs from the target pipeline, the residual geometry will also differ. This means the principal axes might not align, or even if they do, the eigenvalues could vary, either being weaker or stronger, which renders variance-based thresholding less effective. 

The covariance is not scale invariant but is still shift invariant. Hence, we prefer to not use it alone. We propose to also use a distance between residual distributions so that our final loss is sensitive to distribution  shifts. Hence, to be able to learn a relevant pipeline, we consider the following loss function for $\mathtt{TADA}$: 

\begin{equation}
    \mathcal{L} = \lambda
    \underbrace{\|\operatorname{Cov}(\mathcal{R}(\mathcal{S})) - \operatorname{Cov}(\mathcal{R}(\mathcal{T}))\|_F^2}_{\text{\small Geometric alignment}} 
    \  + \ \mu
    \underbrace{\mathcal{D}(\mathcal{R}(\mathcal{S}), \mathcal{R}(\mathcal{T}))}_{\substack{\text{\small Distribution alignment} \\ \text{(e.g., MMD, Wasserstein)}}}
    \text{with } \mu \text{ and } \lambda \text{ to tune}.
\end{equation}

% Expliquer pourquoi les résidus peuvent aider à trouver la pipeline et comment on peut espérer les retrouver.

% Expliquer qu'on peut couper en patchs aligné sur la grille JPEG car l'info importante se trouve dans les patchs 8kx8k

% Proposer nos 2 filtres passe haut en expliquant en quoi ils sont complémentaires. Expliquer que d'autres peuvent être explorés mais que ces 2 là se sont révélés efficaces.

% Expliquer qu'on a intérêt à selectionner les patchs de variance pas trop faible ni trop forte et introduire la notion de threshold.

% Tant qu'à faire, utiliser en source des images RAWs les plus uniformes possibles pour faciliter le calcul.

% Mettre un schéma

%% file: 3_training.tex
\section{Details about $\mathtt{TADA}$ training}\label{sec:trainingdetails}

We detail in this section how we propose to train $\mathtt{TADA}$ to accurately approximate $\omega^{\mathcal T}$.

\subsection{General considerations}

To make noise residuals approximation as accurate as possible, we select randomly 500 RAWs from ALASKABASE \cite{alaskabase} and extract in each of them a 512x512 crop as uniform as possible. This constitutes a RAW base that can be developed for every pipeline to learn. 

% and demosaick them with the amaze algorithm

% At the beginning, all RAWs and JPEG images from targets are rotationed 4 times to artificially augment the dataset. This is possible because orthogonal rotations totally preserve the effect of processing pipelines.

Concerning the development, we initialize it with an identity filter to which we apply a centered gaussian noise with a standard deviation of $0.01$. The constraint of the learnt kernel are artificially enforced at the end of each epoch to avoid disrupting the training phase. Additionally, we do not use a padding strategy for our convolutions, as padding can cause undesirable border effects in the produced images \cite{butora}.

Once RAWs are developed, the differentiable JPEG compression defined in \cite{diffjpeg} is performed, using the quantification table of target images. The initialization choice prevents us from reaching pipelines generating saturated images that ends up to be uniformly black after this JPEG compression.

Right after, we compute noise and target residuals with a combination of two high pass filters to bring more diversity  : the KB filter \cite{kbfilter} and the Laplacian filter with 4 neighbors ($\mathcal L_4$ in \cite{mallet}). 

In order to capture both the intra-bloc and the inter-bloc correlations between JPEG blocs, we compute the covariance matrix from $8\times 16$ patches aligned on the JPEG grid. Such a neighborhood has proven to capture the most relevant information of the development pipeline in ~\cite{taburet}. Within the pool of extracted patches, those exhibiting the lowest variance typically originate from areas of the images with high uniformity. As a result, they fail to effectively differentiate the various emulated pipelines. Indeed, given that kernels summing to 1 are employed, any emulated pipeline yields null residuals within highly uniform region. Conversely, patches with high variance tend to capture textured content where noise residuals are less robust to steganographic schemes. To address this issue, we suggest to select patches with variance falling within the 30th and 60th quantiles of the variance distribution.

The training loss is computed using the selected patches. As explicated in section \ref{sec:analysis}, this loss combines a distance between residuals distributions and a distance between residuals covariances. 
Concerning the distance between distribution, we propose to use the Earth's Mover distance (also called Wassertein $\mathcal W$), a differentiable metric from optimal transport enabling us to avoid vanishing or exploding gradients commonly encountered with other measures like Maximum Mean Discrepancy \cite{feydy} \cite{wassertein}. We also notice that including the Frobenius norm between correlations of residuals enable to speeds up convergence significantly. Smooth learning is enabled by normalizing all our losses with their values at initialization. These normalizations bring all costs to the same scale, making it easier to adjust the learning rate. However, it's more relevant to work with the unnormalized sum of these costs when evaluating any learned pipeline. Keeping the use of normalized costs in the evaluation phase might unfairly favor pipelines that are very good at minimizing one particular cost, while ignoring others. Therefore, at the end of every epoch, $\mathtt{TADA}$ computes the unnormalized sum  $\mathcal{L}_{eval} =
   \|\operatorname{Cov}(\mathcal{R}(\mathcal{S})) - \operatorname{Cov}(\mathcal{R}(\mathcal{T}))\|_F^2
    \ \ + 
    \mathcal{W}(\mathcal{R}(\mathcal{S}), \mathcal{R}(\mathcal{T}))$
and save the final weights minimizing it.

%% file: 4_experiments.tex
\section{Experiments on several targets}\label{sec:experiments}

\subsection{Experimental protocol}

To validate the potential of $\mathtt{TADA}$, we build toy and real-world targets for which we would like to craft tailored sources. The toy targets are created with RawTherapee combining Wavelet Denoising followed by Unsharp Masking and a JPEG compression of QF 85 so that they lead together to high performance drops. It is important to realize that this combination leads to non-linear pipelines. Yet, we will see that $\mathtt{TADA}$ can nevertheless derive meaningful convolutions to generalize on them. The real-world targets are built using the database YFCC100M \cite{yfcc100m} gathering millions of flickr images under CC licenses. From this database, we look for users sharing non-resized images with public quantification tables to comply with our assumptions. This scenario comply with the reality of practitioners since the processing pipeline used by flickr users is totally unknown to us. We finally find 3 users sharing thousands of pictures with the same camera model and compressed with the same quantification tables. Details about all our targets are also presented in Appendix \ref{app:pipelines}.

From these targets, we derive fictitious training and evaluation sets to establish a benchmark for the optimal performance achievable. Additionally, we create an operational set that acts as the  unlabeled small subset available to the forensic analyst useful to train $\mathtt{TADA}$.
All our experiments involves 1.000 cover-stego pairs for training sets, 500 cover-stego pairs for evaluation sets and at maximum 500 images of unknown natures for the operational set. We will consider 3 possibilities for this last set, either it is made of only covers, either it is made of only stegos or either it is made of a balanced mixtures of covers and stegos.  The embedding strategy chosen for all the experiments is UERD \cite{uerd} with a payload of 0.5 bits per non-zero AC DCT coefficient of the luminance channel (bpnzac), a reasonable choice commonly adopted by the steganalysis community \cite{pretrainforstega, wifs2023}. We train $\mathtt{TADA}$ using V100 GPUs. The TIF images to develop are cut into mini-batches of size 256. 

Concerning the RAWs to develop after the learning of a relevant pipeline with $\mathtt{TADA}$, we randomly sample 1.000 textured covers of size $512\times512$ from ALASKABASE using the smart cropping algorithm of the authors \cite{alaskabase}. Then, we develop them with $\omega_{\mathtt{TADA}}$ and create 1.000 cover-stego pairs with UERD to train a steganalysis detector supposed to be efficient on our targets.

For every $\mathtt{TADA}$ learning phase, we fix the maximum number of epochs to 1000. We also select the SGD as our optimizer with a learning rate of 0.001. A learning rate scheduler divides by 2 this learning rate once our $\mathcal L_{eval}$  is not minimized after 100 epochs. At last, an earlystopping procedure stops the learning phase when this same metric did not decrease through 200 successive epochs.

\subsection{Comparison with other strategies}

As our main experiment, we assume to get access to 500 unlabeled images of every target. We propose to learn a relevant source for each of them using 7x7 convolutions with $\mathtt{TADA}$. We then compare the target accuracy obtained with $\mathtt{TADA}$ sources against several methods fighting CSM in a practical scenario where the operational set may be totally unbalanced.
Following the notations of \cite{feasy}, we will name \textit{TgtOnly} the ideal scenario that assume access to labeled target images and, \textit{SrcOnly} the strategy that simply compress RAWs with targets quantification tables.
By using the set-covering strategy of \cite{wifs2022}, we extract eight representatives pipelines $\omega \in \Omega_{R}$ covering the set of 1.000 sources generated by \cite{wifs2023} with a maximum drop of performance of $5\%$. We propose to name \textit{All}, the domain randomization strategy involving to mix these relevant sources  while compressing them with targets quantification tables. Using again $\Omega_R$, we also propose other sota strategies. With the help of a linear multiclassifier, \cite{giboulot} suggests to assign to each target image a specific detector among those trained on $\Omega_R$. Other works such as \cite{csm_fridrich} and \cite{wifs2023} propose to only use the detector trained on the most closest source to the scrutinized target. As explicated by \cite{wifs2023}, this notion of closeness between source and targets could be translated with metrics such as the MMD or the chordal distance (NSCD) between DCTr features \cite{dctr} of each domain. We propose here to add the EarthMover distance and the Frobenius norm between covariance matrices of DCTr, two metrics respectively related to MMD and NSCD that also may be useful.
At last, we suggest to test the subspace alignment of \cite{da2}  and the Covariance Alignment of \cite{da3} since these famous strategies are not impacted by class unbalance. For the subspace alignment, we tune the dimension parameter so that we get the best target accuracy possible.
All the results presented in Table \ref{large-table} are made using a logistic regression on DCTr features of the selected sources.

As demonstrated by this table, $\mathtt{TADA}$ is rather competitive for most of toy strategies while its core assumption of linear processing pipeline is violated. Concerning the realistic targets, its performance are particularly impressive on the SONY target for which we absolute do not know the processing pipeline, highlighting its potential in practical scenarios. Additionally, we notice that the great performance of this strategy are rather stable over our 3 operational balancing, showing its robustness to the embedding scheme. 

\subsection{Limitations}
  We must admit that $\mathtt{TADA}$ is sub-optimal for three targets : RT5, NIKON and CANON. For the RT5 dataset, where the sharpen radius is large, it is probably due to the fact that the convolutional filter is too small. For the NIKON dataset the performances are subpar w.r.t. strategies using the closest cover-set, which relies on a true development pipeline and not and emulated one. For the CANON dataset, non of the proposed approaches are satisfying and $\mathtt{TADA}$, with a poor accuracy of $56\%$ is better than the random strategies with are close to random guesses. We hypothesize  that images from these targets have been developed with different processing pipelines.   

  Following this general experiment, we performed an ablation study modifying several ingredients of $\mathtt{TADA}$ revealing its potential on our real-world targets under different setups. Our results shows that $\mathtt{TADA}$ is still relevant with twice less available target images and can raise the accuracy on SONY and CANON to respectively $77\%$ and $69\%$ if noise residual extractors and training losses are well chosen. We also tested SOTA CNN detectors (e.g. JIN \cite{pretrainforstega}) trained with $\mathtt{TADA}$ sources and we end up with competitive sources against the mixture of the \textit{All} strategy for SONY and NIKON. See more details in Appendix.

\begin{table}[h!]
    \caption{Comparison of $\mathtt{TADA}$ performance vs traditional methods to fight cover source mismatch. The results displayed are target accuracies in \%. 3 cases are studied for the operational set : full cover, balanced mixture of cover and stego, full stego. Best results by target are printed in bold.}
    \label{large-table}
    \centering
    \begin{adjustbox}{max width=\textwidth}
    \begin{tabular}{llllllllll}
      \toprule
        \textbf{Full cover} & RT1 & RT2 & RT3 & RT4 & RT5 & SONY & NIKON & CANON\\
        \midrule
        \rowcolor{mygreen} TgtOnly & 86 & 86 & 86 & 85 & 77  & 92 & 79 & 81 \\
        SrcOnly & 73 & 68 & 68 & 69 & 59 & 54 & 64 & 50 \\
        \midrule
        All  \cite{wifs2022} & 74 & 66 & 68 & 68 & \textbf{68} & 61 & 70 &  52 \\
        \midrule
        Multiclassifier \cite{giboulot} & 67 & 72 & 62 & 62 & 62 & 56 & \textbf{75} & 50 \\
        $\min\limits_{\omega \in \Omega_R} \ \mathrm{NSCD}(\mathrm{DCTr}(\omega),\mathrm{DCTr}(\omega^{\mathcal T}))$ \cite{wifs2023} & \textbf{75} & 76 & 62 & \textbf{77} & 62 & 57 & \textbf{75} & 51 \\
        $\min\limits_{\omega \in \Omega_R}\ ||\mathrm{Cov(\mathrm{DCTr}(\omega))} - \mathrm{Cov(\mathrm{DCTr}(\omega^{\mathcal T}))}||_{F}$ & \textbf{75} & 76 & 62 & \textbf{77} & 62 & 57 & \textbf{75} & 50\\
        $\min\limits_{\omega \in \Omega_R}  \ \mathrm{MMD}(\mathrm{DCTr}(\omega),\mathrm{DCTr}(\omega^{\mathcal T}))$ \cite{wifs2023} & 64 & 76 & 62 & \textbf{77} & 62 & 57 & \textbf{75} & 50 \\
        $\min\limits_{\omega \in \Omega_R} \ \mathcal W(\mathrm{DCTr}(\omega),\mathrm{DCTr}(\omega^{\mathcal T}))$ & 64 & 76 & 62 & \textbf{77} & 62 & 57 & \textbf{75} & 50  \\
        Subspace Alignment \cite{da2} & 72 & 69 & 69 & 70 & 66 & 66 & 51 & 51 \\
        CORAL \cite{da3} & 50 & 50 & 50 & 50 & 50 & 50 & 50 & 50 \\
        $\mathtt{TADA}$ (Ours) & 74 & \textbf{77} & \textbf{77} & 76 & 61 & \textbf{74} & 62 & \textbf{56}  \\
        \bottomrule
    \end{tabular}
    \end{adjustbox}
    \begin{adjustbox}{max width=\textwidth}
    \begin{tabular}{llllllllll}
        \toprule
          \textbf{Mix} &  RT1 & RT2 & RT3 & RT4 & RT5  & SONY & NIKON & CANON\\
          \midrule
          \rowcolor{mygreen} TgtOnly & 86 & 86 & 86 & 85 & 77  & 92 & 79 & 81 \\
          SrcOnly & 73 & 68 & 68 & 69 & 59 & 54 & 64 & 50 \\
          \midrule
          All  \cite{wifs2022} & 74 & 66 & 68 & 68 & \textbf{68} & 61 & 70 &  52 \\
          \midrule
          Multiclassifier \cite{giboulot} & 67 & 72 & 62 & 62 & 62 & 56 & \textbf{75} & 50 \\
          $\min\limits_{\omega \in \Omega_R} \ \mathrm{NSCD}(\mathrm{DCTr}(\omega),\mathrm{DCTr}(\omega^{\mathcal T}))$ \cite{wifs2023}& \textbf{75} & 76 & 62 & \textbf{77} & 62 & 57 & \textbf{75} & 51 \\
          $\min\limits_{\omega \in \Omega_R}\ ||\mathrm{Cov(\mathrm{DCTr}(\omega))} - \mathrm{Cov(\mathrm{DCTr}(\omega^{\mathcal T}))}||_{F}$ & \textbf{75} & 76 & 62 & \textbf{77} & 62 & 57 & \textbf{75} & 50\\
          $\min\limits_{\omega \in \Omega_R}  \ \mathrm{MMD}(\mathrm{DCTr}(\omega),\mathrm{DCTr}(\omega^{\mathcal T}))$ \cite{wifs2023}& 64 & 76 & 62 & \textbf{77} & 62 & 57 & \textbf{75} & 50 \\
          $\min\limits_{\omega \in \Omega_R} \ \mathcal W(\mathrm{DCTr}(\omega),\mathrm{DCTr}(\omega^{\mathcal T}))$ & 64 & 76 & 62 & \textbf{77} & 62 & 57 & \textbf{75} & 50  \\
          Subspace Alignment \cite{da2} & 72 & 70 & 69 & 69 & 66 & 66 & 51 & 51 \\
          CORAL \cite{da3} & 50 & 50 & 50 & 50 & 50 & 50 & 50 & 50 \\
          $\mathtt{TADA}$ (Ours) & 74 & \textbf{77} & \textbf{77} & 76 & 63 & \textbf{68} & 62 & \textbf{56}  \\
          \bottomrule
    \end{tabular}
    \end{adjustbox}
    \begin{adjustbox}{max width=\textwidth}
    \begin{tabular}{lllllllll}
            \toprule
            \textbf{Full stego} &  RT1 & RT2 & RT3 & RT4 & RT5 & SONY & NIKON & CANON\\
            \midrule
            \rowcolor{mygreen} TgtOnly & 86 & 86 & 86 & 85 & 77  & 92 & 79 & 81 \\
            SrcOnly & 73 & 68 & 68 & 69 & 59 & 54 & 64 & 50 \\
            \midrule
            All  \cite{wifs2022} & 74 & 66 & 68 & 68 & \textbf{68} & 61 & 70 &  52 \\
            \midrule
            Multiclassifier \cite{giboulot} & 67 & 72 & 62 & 62 & 62 & 56 & \textbf{75} & 50 \\
            $\min\limits_{\omega \in \Omega_R} \ \mathrm{NSCD}(\mathrm{DCTr}(\omega),\mathrm{DCTr}(\omega^{\mathcal T}))$ \cite{wifs2023} & \textbf{75} & 76 & 62 & \textbf{77} & 62 & 57 & \textbf{75} & 51 \\
            $\min\limits_{\omega \in \Omega_R}\ ||\mathrm{Cov(\mathrm{DCTr}(\omega))} - \mathrm{Cov(\mathrm{DCTr}(\omega^{\mathcal T}))}||_{F}$ & \textbf{75} & 76 & 62 & \textbf{77} & 62 & 57 & \textbf{75} & 50\\
            $\min\limits_{\omega \in \Omega_R}  \ \mathrm{MMD}(\mathrm{DCTr}(\omega),\mathrm{DCTr}(\omega^{\mathcal T}))$ \cite{wifs2023} & 64 & 76 & 62 & \textbf{77} & 62 & 57 & \textbf{75} & 50 \\
            $\min\limits_{\omega \in \Omega_R} \ \mathcal W(\mathrm{DCTr}(\omega),\mathrm{DCTr}(\omega^{\mathcal T}))$ & 64 & 76 & 62 & \textbf{77} & 62 & 57 & \textbf{75} & 50  \\
            Subspace Alignment \cite{da2} & 70 & 70 & 70 & 69 & 66 & 66 & 51 & 51 \\
            CORAL \cite{da3} & 50 & 50 & 50 & 50 & 50 & 50 & 50 & 50 \\
            $\mathtt{TADA}$ (Ours) & \textbf{75} & \textbf{77} & \textbf{77} & 75 & 62 & \textbf{70} & 62 & \textbf{53}  \\
            \bottomrule
      \end{tabular}
      \end{adjustbox}
    \end{table}

  %RT5 is the target issued from the strongest processing pipeline among our toy targets. Hence, this is not surprising that our simple linear approximation fail for this specific case. Concerning Nikon and Canon, we do not know the processing pipelines involved but we suspect them to be highly non linear explaining our failures. Since CANON is really hard to learn, we believe that images from these targets have been developed with different processing pipelines, explaining why it is not possible for TADA to learn one supposed to derive all.

\section{Conclusion}
To tackle Cover Source Mismatch in steganalysis, we introduce $\mathtt{TADA}$, a strategy enabling to create tailored sources for any steganalysis target even when they are small and highly unbalanced. This strategy lies on the alignment of noise residuals in terms of geometry (with their covariance) and distributions (with a Wassertein metric).
Several experiments are conducted to demonstrate that our strategy is promising and can outperform traditional methods fighting CSM in practical scenarios. However, we also observe that $\mathtt{TADA}$'s power is limited by its oversimplistic assumption that every pipeline can be approximated with only one convolution. Despite this limitations, there is considerable room for improvement considering the efficiency of neural networks to handle non-linear relationships.
Furthermore, in scenarios where target images result from different processing pipelines, $\mathtt{TADA}$, by its construction, cannot perform effectively. Nevertheless, with a clustering approach harnessing the features of \cite{mallet}, we can try to group images based on their processing pipeline proximity, offering a potential solution by applying $\mathtt{TADA}$ to each cluster. To conclude, $\mathtt{TADA}$ pave the way for more robust and reliable steganalysis in diverse and complex real-world environments.

\section*{Acknowledgements}

Our experiments were possible thanks to computing means of IDRIS through the resource allocation 2023-AD011013285R2 assigned by GENCI. This work received funding from the European Union’s
Horizon 2020 research and innovation program under grant agreement No 101021687 (project “UNCOVER”) 
and the French Defense \& Innovation Agency. The work of Tomáš Pevný  was supported by Czech Ministry of Education 19-29680L.

\clearpage

%% file: 5_appendix.tex
\clearpage

\section{Appendix / supplemental material}

\subsection{Details abouts the different sources used in our experiments.}

\label{app:pipelines}

\begin{table}[h]
    \caption{Details about parameters of the processing pipelines generating toy targets}
    \centering
    \begin{tabular}{ccccc}
        \toprule
        \textbf{Toy Targets} & \textbf{Denoise Luma} & \textbf{Sharpen Radius} & \textbf{Sharpen Amount} &  \textbf{Sharpen Thresholds} \\
        \midrule
        RT1 & 26 & 0.176 & 225 & 20;80;2000;1200; \\
        RT2 & 30 & 0.01& 225& 20;80;2000;1200;   \\
        RT3 & 30 & 0.176& 225 & 20;80;2000;1200; \\
        RT4 & 30 & 0.341 & 225& 20;80;2000;1200;  \\
        RT5 & 30 & 1.5 & 225 & 20;80;2000;1200; \\
        \bottomrule
    \end{tabular}
    \label{tab:toy}
  \end{table}
  
  \begin{table}[h]
    \caption{Details about real world targets}
    \centering
    \begin{tabular}{cccc}
        \toprule
        \textbf{Real-world Targets} & \textbf{Username} & \textbf{Camera Model} & \textbf{Quality Factor} \\
        \midrule
        SONY & toms travels & SONY SLT A37 & 90  \\
        CANON & Andy E. Nystrom & Canon PowerShot SX30 IS & 93   \\
        NIKON & NR Acampamentos & NIKON D40 & 90 \\
        \bottomrule
    \end{tabular}
    \label{tab:rw}
  \end{table}
  
\subsection{CSM interpretation}

% \label{appendix:a}

In this section, we complete our geometric interpretation of CSM considering scatter plots of neighboring residuals. We propose to illustrate why it's important to equalize spread (eigenvalues) on top of principal directions (eigenvectors). To do so, we propose to consider the 3x3 sharpening filter $\mathcal S$ : 

\begin{equation}
    \label{eq:sharpen}
    \setlength{\arraycolsep}{10pt} % Adjust the value to increase space between columns
    \renewcommand{\arraystretch}{2} % Adjust the value to increase space between rows
    \begin{bmatrix}
      0 & - \dfrac{1}{4}  & 0 \\
      -\dfrac{1}{4} & 2 & -\dfrac{1}{4}  \\
      0 & -\dfrac{1}{4}  & 0
    \end{bmatrix}
\end{equation}

followed by a JPEG compression of QF 100.

% And the denoising filter $\mathcal D$ :

% \begin{equation}
%     \label{eq:sharpen}
%     \setlength{\arraycolsep}{15pt} % Adjust the value to increase space between columns
%     \renewcommand{\arraystretch}{3} % Adjust the value to increase space between rows
%     \begin{bmatrix}
%       \dfrac{1}{9} & \dfrac{1}{9} & \dfrac{1}{9} \\
%       \dfrac{1}{9} & \dfrac{1}{9} & \dfrac{1}{9} \\
%       \dfrac{1}{9} & \dfrac{1}{9} & \dfrac{1}{9}
%     \end{bmatrix}
% \end{equation}

% Figure~\ref{fig:directions} illustrates the fact that the 

% \begin{figure}[h]
%     \centerline{
%     \includegraphics[width=\linewidth]{diff_directions.pdf}}
%     \caption{Geometric interpretation of CSM originating from two different processing operations : sharpening (left) and denoising (right). Each point represents two neighboring samples of an image subject to a high-pass filter, {\it i.e.} a 2D residual. %The stego residuals have a slightly higher variance in the direction of the principal axis, a difference that can be used to separate cover and stego of a same source projecting on this axis. 
%     %The principal axis being different in both cases, projecting residuals with a wrong principal axis leads to difficulties to separate cover and stegos.
%     The two distributions are rather different w.r.t. their scales and orientations, just leading to covariate shift.
%     }
%     \label{fig:directions}
% \end{figure}

Then, we propose to compare the scatterplots of neighboring samples from 2D residuals with the same filter multiplied by a 0.5 coefficient.

\begin{figure}[h!]
    \centerline{
    \includegraphics[width=0.7\linewidth]{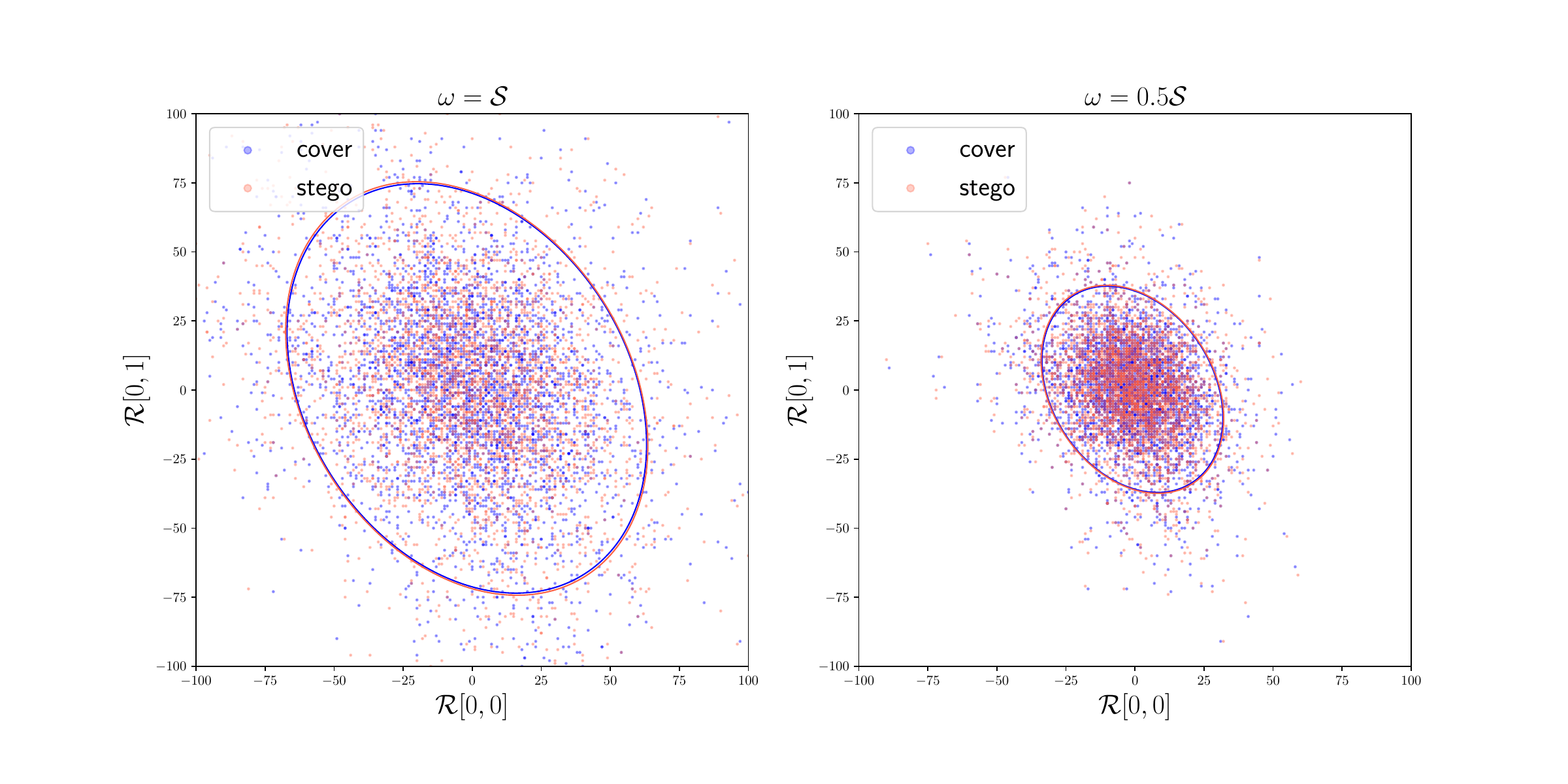}}
    \caption{Geometric interpretation of CSM originating from processing operations differing from a multiplicative factor  : $\mathcal S$ (left) and 0.5$\mathcal S$ (right). Each point represents two neighboring samples of an image subject to a high-pass filter, {\it i.e.} a 2D residual.}
    \label{fig:variance}
\end{figure}

On Figure \ref{fig:variance} stego residuals have a slightly higher variance in the direction of the principal axis, a difference that can be used to separate cover and stego of a same source projecting on this axis. The principal axis is similar in both cases but, the threshold used to separate cover and stego is not the same potentially leading to CSM. 

% %Preciser que la correlation est identique dans ce cas par scale invariance alors qu'il y a un bien un regret !!

In this case, since JPEG QF is 100, the images resulting from 0.5$\mathcal S$ are just resulting from a 0.5 multiplication of every pixel from $\mathcal S$, hence leading to identical residual correlations. However, a gap of performance is observed when we train on one source and evaluated on the other as highlighted in \ref{tab:sharpen}.

\begin{table}[h!]
    \caption{Accuracy matrix showing the mismatch between $\mathcal S$ and 0.5$\mathcal S$ . 2000 images for train and 1000 for test, the embedding strategy is UERD \cite{uerd} with a payload of 0.5bpnzac.}
    \label{tab:sharpen}
    \centering
    \begin{tabular}{lccc}
        \toprule
        \multicolumn{2}{l}{\textit{Train      \textbackslash \ Test}} &  $\mathcal S$ & $0.5\mathcal S$ \\
        \midrule
        $\mathcal S$ &  &65 & 61\\
        $0.5\mathcal S$ & &  64 & 70 \\
        \bottomrule
    \end{tabular}
\end{table}

\subsection{Ablation studies}

\label{app:ablations}

We perform several ablation studies to assess all the potential and limitations of $\mathtt{TADA}$ for real-world targets. For each study of this type, we propose to start from the main study of the paper in section \ref{sec:experiments} while changing only one ingredient at a time. 

From Table \ref{a1}, it seems that cutting patches of size $16\times16$ enables to enhance $\mathtt{TADA}$ performance. This is not a complete surprise as raising the size of the patch means providing more information about noise residuals distributions.

\begin{table}[h]
    \caption{Ablation I : What happen if we extract residual patches of different size ?}
    \centering
    \begin{tabular}{lcccc}
        \toprule
        \multicolumn{2}{l}{\textit{Target      \textbackslash \ Patch size}} & $8 \times 8$ & $8 \times 16$ & $16 \times 16$ \\
        \midrule
        SONY &  & 75 & 70 & 75 \\
        NIKON  &  & 63 & 62 & 62 \\
        CANON  &  & 53 & 53 & 57 \\
        \bottomrule
    \end{tabular}
    \label{a1}
\end{table}

Table \ref{a2} highlights that raising the kernel size is not a solution enabling to perform well on any target. We see for instance that leveraging a small kernel enables to better generalize on NIKON while a rather large kernel is more interesting for SONY. We understand from this study that it's important to inject in $\mathtt{TADA}$, a convolution as close as possible to the true operation applied to target pipelines. 

\begin{table}[h]
    \caption{Ablation II: What happen if the kernel size change ?}
    \centering
    \begin{tabular}{lcccccc}
        \toprule
        \multicolumn{2}{l}{\textit{Target \textbackslash Kernel size}} & $3 \times 3$ & $5 \times 5$ & $7 \times 7$ & $9 \times 9$ & $11 \times 11$ \\
        \midrule
        SONY &  & 52 & 67 & 70 & 74 & 72 \\
        NIKON &  & 61 & 64 & 62 & 59 & 60 \\
        CANON &  & 50 & 60 & 53 & 61 & 60 \\
        \bottomrule
    \end{tabular}
    \label{a2}
\end{table}

Table \ref{a3} shows the robustness of $\mathtt{TADA}$ when only few available samples from the targets is available. We see for instance that the generalization on SONY is pretty impressive in all the cases studied. Moreover, it seems that the performance on CANON is even slightly better using  smaller sets of images from this target. We think this results is coming from our patch selection. With very few images, more diverse patches are taken into account even though they present very low or very high variances, and that may be slightly beneficial for CANON. This is also consistent with Table \ref{a7}.

\begin{table}[h!]
    \caption{Ablation III: What happen if the operational set is smaller ?}
    \centering
    \begin{tabular}{lccccc}
        \toprule
        \multicolumn{2}{l}{\textit{Target \textbackslash  \ Amount of operational data}} & $10$ & $100$ & $200$ & $500$ \\
       \midrule
        SONY &  & 66 & 71 & 70 & 70 \\
        NIKON &  & 62 & 63 & 62 & 62  \\
        CANON &  & 56 & 57 & 57 & 53 \\
        \bottomrule
    \end{tabular}
    \label{a3}
\end{table}

Table \ref{a4} shows the complementarity of our training losses. We can see for instance that combining the three losses is the best strategy to generalize on SONY. However, we succeed to achieve a very great performance on CANON using only the Wassertein, and a competitive performance on NIKON combining the Wassertein and the Frobenius between correlations. This study invites us to rethink our training normalization so that we can harness the best from each loss.

\begin{table}[h!]
    \caption{Ablation IV: What happen if we change the training loss ?}
    \centering
    \begin{tabular}{lcccccccc}
        \toprule
        \multicolumn{2}{l}{\textit{Target \textbackslash \ Patch size}} & $\mathcal{W}$ & $\mathcal{L}_{corr}$ & $\mathcal{L}_{cov}$& $\mathcal{W}  + \mathcal{L}_{cov}$ &  $\mathcal{W}  + \mathcal{L}_{corr}$& $\mathcal{L}_{cov} + \mathcal{L}_{corr}$ & $\mathcal{W} + \mathcal{L}_{cov} + \mathcal{L}_{corr}$  \\
        \midrule
        SONY &  & 52 & 51 & 61 & 60 & 53 & 67 & 70\\
        NIKON &  & 54 & 65 & 57 & 55 & 68 & 63 & 62\\
        CANON &  & 69 & 60 & 59 & 56 & 59 & 57 & 53\\
        \bottomrule
    \end{tabular}
    \label{a4}
\end{table}

Table \ref{a5} shows that the choice of the residual extractor is crucial for $\mathtt{TADA}$. For instance, using only the KB filter, we jump from an accuracy of 70\% on SONY to an accuracy of 77\%, an impressive result for a source for which we ignore the processing pipeline. In the same vein, we are much better on CANON with the KB Filter.

\begin{table}[h!]
    \caption{Ablation V: What happen if we change our residual extractor ?}
    \centering
    \begin{tabular}{lccccc}
        \toprule
        \multicolumn{2}{l}{\textit{Target \textbackslash  \ Residual extractor}} & $\mathcal{L}_4$ \cite{mallet}  & KB  \cite{kbfilter} & $\mathcal{L}_4$ and KB\\
       \midrule
        SONY &  & 66 & 77 & 70\\
        NIKON &  & 52 & 57 &  62\\
        CANON &  & 58 & 60 & 53\\
        \bottomrule
    \end{tabular}
    \label{a5}
\end{table}

Table \ref{a6} shows that our kernel constraint can be relaxed without drop of performance on our three targets. It suggests to only keep the symmetry constraint and forget the normalization to 1. 

\begin{table}[h!]
    \caption{Ablation VI: What happen if the we relax the kernel constraint ?}
    \centering
    \begin{tabular}{lccccc}
        \toprule
        \multicolumn{2}{l}{\textit{Target \textbackslash  \ Constraint}} & None & Sum to 1 & Symmetry & Sum to 1 and Symmetry \\
       \midrule
        SONY &  & 73 & 70 & 72 & 70 \\
        NIKON &  & 63 & 64 & 64 & 62  \\
        CANON &  & 58 & 56 & 59 & 53 \\
        \bottomrule
    \end{tabular}
    \label{a6}
\end{table}

Finally, Table \ref{a7} suggests that our patch selection is very relevant for SONY but detrimental for NIKON and CANON. This invite us to rethink our patch selections so that we can perform the best on any target.

\begin{table}[h!]
    \caption{Ablation VI : What happen if we relax the patch selection ?}
    \label{a7}
    \centering
    \begin{tabular}{lccc}
        \toprule
        \multicolumn{2}{l}{\textit{Source      \textbackslash \ Target}} &  Without patch selection & We patch selection \\
        \midrule
        SONY &  &54 & 70\\
        NIKON & & 66 & 62 \\
        CANON & & 55 & 53 \\
        \bottomrule
    \end{tabular}
\end{table}

\clearpage

\subsection{Results with deep neural networks}

We present at last some results we got using SOTA deep detectors with the source learnt by $\texttt{TADA}$ during the main experiment of the paper. This time, $\texttt{TADA}$ is competitive against SrcOnly and All for SONY and NIKON targets. This is particularly interesting considering that it was previously hard to generalize on NIKON using linear classifiers trained on $\texttt{TADA}$ sources. Moreover, we again observe that it's very difficult to generalize on CANON, even though we saw in Table  \ref{a4} that it's possible to reach a competitive accuracy of 69\% with a $\texttt{TADA}$ source, only using the Wassertein to learn it. Finally, we observe again the robustness of $\texttt{TADA}$ against highly unbalanced targets.

\begin{table}[h]
  \caption{Target accuracies using JIN\cite{pretrainforstega} on different sources}
  \centering
  \begin{tabular}{ccccccccc}
      \toprule
      \textbf{Strategies} & \textbf{RT1} & \textbf{RT2} & \textbf{RT3} & \textbf{RT4} & \textbf{RT5} & \textbf{SONY} & \textbf{NIKON} & \textbf{CANON} \\
      \midrule
      \rowcolor{mygreen} TgtOnly & 91 & 94 & 93 & 94 & 92 & 87 & 93 & 91 \\
      SrcOnly & 85 & 84 & 84  & 84  & \textbf{80}  & 81 & 90 & \textbf{63} \\
      \midrule
      All & \textbf{91} & \textbf{89} & \textbf{90} & \textbf{91} & 77 & 75 & 93 & 53 \\
      \midrule
      TADA (Full Cover) & 86 & 83 & \textbf{90} & 85 & 65 & \textbf{85} & \textbf{96} & 54 \\
      TADA (Mix) & 85 & 84 & 88 & 87 & 66 & \textbf{84} & \textbf{96} & 55 \\
      TADA (Full Stego) & 84 & 87 & 87 & 84 & 70 & \textbf{85} & \textbf{96} & 58 \\
      \bottomrule
  \end{tabular}
  \label{tab:devices}
\end{table}